\documentclass[amssymb,11pt]{article}
\usepackage[pdftex,colorlinks=true,urlcolor=black,filecolor=black,linkcolor=black,
            pdftitle={Fractional Derivative Cosmology.},
            pdfauthor={Mark D. Roberts},
            pdfsubject={Fractional derivatives, field equations},
            pdfkeywords={fractional derivative, field equations},
            pagebackref,pdfpagemode=None,bookmarksopen=true]{hyperref}
\usepackage{amssymb}
\usepackage{graphicx}
\newcommand{\bc}{\begin{center}}
\newcommand{\ec}{\end{center}}
\newcommand{\be}{\begin{equation}}
\newcommand{\ee}{\end{equation}}
\newcommand{\ber}{\begin{eqnarray}}
\newcommand{\ear}{\end{eqnarray}}
\newcommand{\Lg}{{\cal L}}
\newcommand{\n}{\nonumber\\}

\newcommand{\Op}{{\cal D}}

\newcommand{\pr}{{\cal P}}

\newcommand{\tf}{\textstyle\frac}

\newcommand{\p}{\partial}
\begin{document}
%%%%%%%%%%%%%%%%%%%%%%%%%%%%%%%%%%%%%%%%%%%%%%%%
\title{Fractional Derivative Cosmology.}
\author{
\href{http://www.violinist.com/directory/bio.cfm?member=robemark}
{Mark D. Roberts},\\
54 Grantley Avenue,  Wonersh Park,  GU5 0QN,  UK\\
mdr@ihes.fr
}
\date{$6^{th}$ of September 2009}
\maketitle
%%%%%%%%%%%%%%%%%%%%%%%%%%%%%%%%%%%%%%%%%%%%%%%%%%%%%%%%%%%%%%%%%%%%%%%%%%%%%%%%%%%%%%%%%%%%%
\begin{abstract}
%%%%%%%%%%%%%%%%%%%%%%%%%%%%%%%%%%%%%%%%%%%%%%%%%%%%%%%%%%%%%%%%%%%%%%%%%%%%%%%%%%%%%%%%%%%%%
The degree by which a function can be differentiated need not be restricted to integer values.
Usually most of the field equations of physics are taken to be second order,
curiosity asks what happens if this is only approximately the case and the field equations are nearly second order.
For Robertson-Walker cosmology there is a simple fractional modification of the
Friedman and conservation equations.
In general fractional gravitational equations similar to Einstein's are hard to define
as this requires fractional derivative geometry.
What fractional derivative geometry might entail is briefly looked at and it turns out that
even asking very simple questions in two dimensions leads to ambiguous or intractable results.
A two dimensional line element which depends on the Gamma-function is looked at.
\end{abstract}%%%%%%%%%%%%%%%%%%%%%%%%%%%%%%%%%%%%%%%%%%%%%%%%%%%%%%%%%%%%%%%%%%%%%%%%%%%%%%%
{\tiny\tableofcontents}
\section{Introduction}\label{intro}%%%%%%%%%%%%%%%%%%%%%%%%%%%%%%%%%%%%%%%%%%%%%%%%%%%%%%%%%%
\subsection{Motivation}\label{motivation}%%%%%%%%%%%%%%%%%%%%%%%%%%%%%%%%%%%%%%%%%%%%%%%%%%%%%
The field equations of fundamental physics are usually taken to be second order.
Like any other property of a physical theory this should be subjected
to experimental and observational tests to see what the experimental bounds are.
A method of examining this is by using fractional derivatives to investigate the
properties of differential equations which are almost second order.
Electromagnetism might produce the best tests of how near to second order governing
field equations need to be,  but here an attempt is made to see what modification of
gravity theory can be constructed using fractional derivatives.
\subsection{History}\label{history}%%%%%%%%%%%%%%%%%%%%%%%%%%%%%%%%%%%%%%%%%%%%%%%%%%%%%%%%%%
In 1695 L'H\^opital wrote to Leibniz asking him what happened if the number of times a function
was differentiated was not an integer but rather $\tf{1}{2}$,
from this early beginning the subject of fractional derivatives was born.
Although the derivatives are called fractional derivatives
they can take any real or sometimes complex value.
Recent textbooks include \cite{KST,podlubny}.
Recent applications to physics include construction of a fraction Schr\"odinger equation
\cite{naber,laskin} and some properties of field theories \cite{BM,zavada}.
\subsection{Methodology}\label{methodology}%%%%%%%%%%%%%%%%%%%%%%%%%%%%%%%%%%%%%%%%%%%%%%%%%%%%%
There are two distinct methods of approaching what fractional derivative cosmology could be.
The simplest is {\em last step modification} in which Einstein's field equations for a given
geometric configuration are replaced with analogous fractional field equations,
in other words $\p_a\rightarrow\Op_a^k$ after the field equations for a specific geometry have been derived.
The fundamentalist methodology is {\em first step modification} in which one starts by constructing
fractional derivative geometry.
The problem with the last step approach is that it probably only gives consistent answers
for geometric configurations that are expressed in rectilinear coordinates;
the problem with the first step approach is that the whole of geometry has to be rethought through,
even things like linear coordinate transformations have to be replaced by fractional ones,
perhaps quadratic forms by fractional forms and so on;
intermediate step appraoches seem to have the disadvantages of both of the above approaches.
When using fractional derivatives it is not always the case that differentiating
a constant gives zero see Podlubny \cite{podlubny}p.80,
so it is not clear what kind of fractional derivative to use,
here a pragmatic approach is adopted:
if a fractional derivative can be made to work then it is used.
\subsection{Outline \& Conventions}\label{outline}%%%%%%%%%%%%%%%%%%%%%%%%%%%%%%%%%%%%%%%%%%%%%%%%%%%%%%%%%%
In \S\ref{newt} an attempt is made to see how the newtonian $\tf{1}{r}$ potential is modified when
the d'Almbertian is replaced at the 'last step' by fractional derivatives,  it is found that the
'last step' assumption is not necessarily correct as spherically flat spacetime need not take its
familiar form.
In \S\ref{rw} Robertson-Walker cosmology is looked at in the last step approach.
In \S\ref{sfs} briefly looks at what fractional derivative geometry might entail.
In \S\ref{conc} is the conclusion.
p is used for degree of fractional differentiation,
the familiar non-fractional differential equations are recovered when $p=1$,
$\pr,\mu,U_a$ are the pressure, density and co-moving vector of a fluid.
Calculations were carried out using maple9/grtensorII \cite{MPL}.
\section{Newtonian Gravity}\label{newt}%%%%%%%%%%%%%%%%%%%%%%%%%%%%%%%%%%%%%%%%%%%%%%%%%%%%%%%%
Vacuum Newtonian gravity is taken to be governed by the D'Almbertian operator acting on a scalar function
\be
\Box\phi=0,
\label{dalmbertian}
\ee
Applying the last step fractional derivative substitution approach for spherical symmetry gives
\be
\Op^{2p}\psi+\frac{2}{r}\Op^p\psi=0
\label{lss}
\ee
numerically solving this for values close to one produces potentials as illustrated in figure one.
%\begin{figure}
%\includegraphics[height=5in]{7april09.pdf}
%%\includegraphics[height=3.5in,angle=90]{7april09.pdf}
%%{\epsfig{figure=7april09.eps,height=3in}}
%\caption{lowest line$=2.10^4\exp(-r^{0.1}/0.1)$,
%         middle line$=1/r$,
%         top line$=5.10^{-5}\exp(r^{-0.1}/0.1)$}
%\label{FigureOne}
%\end{figure}
For $p$ close to one the potential can be made arbitarly close to the non-fractional case.
Rates of decay of potentials have been discussed at lenght in \cite{mdr39}.
At first sight one might imagine that all that now needs doing is using either expansions or numerical analysis
to compare this result with observations:  but there is there is a serious problem with equation \ref{lss},
the $\tf{2}{r}$ term assumes a standard spherically symmetric flat background spacetime,  however for fractional
derivatives this assumption does not necessarily hold,  the two-sphere in spherical coordinates
is not necessarily the familiar one as this requires assumptions of standard geometry
which might not hold in fractional geometry,  see \S\ref{ffs}.
\section{Fractional Robertson-Walker Cosmology}\label{rw}%%%%%%%%%%%%%%%%%%%%%%%%%%%%%%%%%%%%%
\subsection{Friedman Equation}\label{feq}%%%%%%%%%%%%%%%%%%%%%%%%%%%%%%%%%%%%%%%%%%%%%%%%%%%%%%
For the moment we retain the standard quadratic form of the Robertson-Walker line element
\be
ds^2=-dt^2+\frac{a(t)^2}{(1+k(x^2+y^2+z^2)/4)^2}\left(dx^2+dy^2+dz^2\right),
\label{rwqle}
\ee
where $k=0,\pm1$.
The stress is taken to be that of a perfect fluid
\be
T_{ab}=(\mu+\pr)U_aU_b+\pr g_{ab},~~~
U_a=\left(1,0,0,0\right),
\label{pf}
\ee
which for present purposes contains no derivatives,  this might change if the Clebsch
representation of the comoving vector field $U_a$ is used.
For general relative the dynamics are governed by the Friedman equation
and the first conservation equation $U_aT^{ab}_{..;b}=0$ replacing partial derivatives
with fractional derivatives at the last moment the Friedman and conservation equations become
\be
3\left[k+\left(\Op^p_ta\right)^2\right]=\kappa\mu a^2,~~~~~~~
a^3\Op^p_t\pr=\Op^p_t\left[a^3\left(\mu+\pr\right)\right],
\label{ffe}
\ee
respectively.
For the simpest non-fraction $p=1$ case take both $k,\pr=0$
then the conservation equation integrates to
\be
\mu=\frac{C}{a^3},
\label{firstint}
\ee
and then the Friedman equation integrates to give
\be
a=\left(\frac{\kappa C}{3}t\right)^\frac{2}{3}.
\label{simplest}
\ee
For the $k\mp 1$ cases powers of $t$ are replaced be trigonometric or hyperbolic functions.
\subsection{Naive Approach}\label{na}%%%%%%%%%%%%%%%%%%%%%%%%%%%%%%%%%%%%%%%%%%%%%%%%%%%%%
A naive generalization of the above non-fractional approach is just to replace the time
derivative in the Friedman equation with a fractional time derivative.
The Riemann-Liouville fractional derivative with lower terminal at 0,
after correcting a typo \cite{podlubny} page 310 is
\be
\Op_t^pt^\nu=\frac{\Gamma(\nu+1)}{\Gamma(\nu+1-p)}t^{\nu-p},
\label{fracdevpower}
\ee
Absorbing constants into $C$ gives the scale factor and density
\be
a=Ct^\frac{2p}{3},~~~
\mu=C^{-2}t^{-2p},
\label{rwsimple}
\ee
which is what might have been anticipated.
There is a problem with the above in integrating the conservation equation \ref{firstint} it was assumed
that differentiating a constant gives zero but \ref{fracdevpower} gives
\be
\Op_t^pB=\frac{B}{\Gamma(1-p)}t^{-p},
\label{cefr}
\ee
which is non-zero for non-zero $B$.
\subsection{Compensating Pressure.}\label{cs}%%%%%%%%%%%%%%%%%%%%%%%%%%%%%%%%%%%%%%%%%%%%%%%
To rectify this a compensating pressure can be used,
substituting \ref{rwsimple} back in to \ref{ffe}
a pressure which satisfies the conservation equation is
\be
\pr=At^{-2p},~~~~~~~
A^{-1}C^{-2}=\frac{\Gamma(1-p)\Gamma(1-2p)}{\Gamma(1-3p)}-1,
\label{cpres}
\ee
$AC^2$ is plotted in figure two.
%\begin{figure}
%\includegraphics[height=5in]{compressure.pdf}
%%\includegraphics[height=3.5in,angle=90]{7april09.pdf}
%%{\epsfig{figure=7april09.eps,height=3in}}
%\caption{Red line is $AC^2$,  blue points are $y=0.35(x-1)$}
%\label{Compensating Pressure}
%\end{figure}
This shows that the compensating pressure is smaller than the density.
\subsection{$\gamma$-equation of state.}\label{ges}%%%%%%%%%%%%%%%%%%%%%%%%%%%%%%%%%%%%%%%%%%%%%
The solution \ref{cpres} is an example of a $\gamma$-equation of state for which
\be
p=(\gamma-1)\mu,
\label{defg}
\ee
for \ref{cpres} $\gamma=AC^2+1$.
To investigate if \ref{cpres} can be generalized choose \ref{defg} and
\be
a=At^m,~~~\mu=Ct^n,
\label{gant}
\ee
then the Friedman equation restricts the values of $n$ and $C$
\be
n=-2p,~~~C=\frac{3}{\kappa}\frac{\Gamma(m+1)^2}{\Gamma(m+1-p)^2},
\label{fgam}
\ee
and the conservation equation places the restriction
\be
\frac{(\gamma-1)}{\gamma}\frac{\Gamma(1-2p)}{\Gamma(1-3p)}
=\frac{\Gamma(3m-2p+1)}{\Gamma(3m-3p+1)},
\label{egam}
\ee
because $m$ occurs inside a $\Gamma$-function it is hard to access the use of this solution.
Generalizing \ref{gant} with $\pr=Bt^r$ the conservation equation gives $n=r$ so that the results are the same.
\section{Fractional Derivative Geometry}\label{sfs}%%%%%%%%%%%%%%%%%%%%%%%%%%%%%%%%%%%%%%%%%%%%%%%%%%%%%%%%%%
\subsection{Fractional alterations}
A generalized Christoffel symbol can be taken to be
\be
\Gamma^a_{bc}=\frac{1}{2}g^{ad}\left(\Op_cg_{bd}+\Op_bg_{cd}-\Op^dg_{bc}\right),
\label{genchristoffel}
\ee
where $\Op_c$ is an operator with one spacetime index,  in particular it can be a vector field,
or the familiar partial differential operator,  or a fractional derivative.
A generalized Riemann tensor can be taken to be
\be
GR^a_{bcd}=\Op_c\Gamma^a_{db}-\Op_d\Gamma^a_{cb}+\Gamma^a_{cf}\Gamma^f_{db}-\Gamma^a_{df}\Gamma^f_{cb},
\label{genriemann}
\ee
where for simplicity the operator $\Op_c$ is chosen to be the same one as in the generalized
christoffel symbol.
The fractional quadratic form could be
\be
ds^{2p}=g_{ij}\left(dx^i\right)^p\left(dx^j\right)^p,
\label{fracform}
\ee
for $p=Z_+/2$ the metric requires $2\times\alpha$ indices,
otherwise the number of indices is not integer,
the meaning of a metric with non-integer number of indices is not clear.
Forms of the form \ref{fracform} can occur in Finsler geometry \cite{chern},
however in Finsler geometry there are still linear coordinate transformations
rather than fractional ones of the form \ref{fraccoordtrans}.
\subsection{Fractional Flat Space}\label{ffs}%%%%%%%%%%%%%%%%%%%%%%%%%%%%%%%%%%%%%%%%%%%%%%%%
It is now possible to ask what is fractional rectilinear flat spacetime,
for simplicity working in two dimensions with the usual quadratic form
\be
ds^2=-N(t,x)^2dt^2+F(t,x)^2dx^2,
\label{fracflat}
\ee
if the generalized Christoffel symbol \ref{genchristoffel} vanishes
then the generalized Riemann tensor \ref{genriemann} will also vanish,
looking at one component of the generalized Christoffel symbol
\be
\Gamma^t_{tt}=\frac{1}{2N^2}\Op_tN^2
\label{gamt}
\ee
this will not vanish unless either $N=0$ in which case \ref{fracflat} degenerates or
caputso derivatives are used which results in the familar two dimensional rectilinear flat
space with $N=F=1$.  The next question is what is fractional spherical flat spacetime.
In two dimensions this question is what is the analog of the following:  start with line element
\be
ds^2=dx^2+dy^2,
\label{standardle}
\ee
and then perform the transformation
\be
x=r\sin(\theta),~~~y=r\cos(\theta),
\label{standardct}
\ee
to give the line element
\be
ds^2=dr^2+r^2d\theta^2.
\label{standardss}
\ee
For the fractional case start with \ref{standardle} and use coordinate transformations
\be
dx^i=\Op^p_ix^idx^{i`},
\label{fraccoordtrans}
\ee
choosing \ref{standardct} to be replaced by
\be
x=r^p\sin(\theta),~~~y=r^p\cos(\theta),
\label{fracct}
\ee
gives the line element
\be
ds^2=\Gamma(p+1)^2dr^2+2\Gamma(p+1)r^p\cos(\frac{\pi p}{2})drd\theta+r^{2p}d\theta^2.
\label{newd2}
\ee
Altering the power of the trignometric parts of \ref{fracct} does not seem to remove the cross
term but rather makes things more complicated.  Maple/grtensorII \cite{MPL} works out the standard,
non-fractional derivative curvature to be
\ber
R_{r\theta r\theta}=-p(p-1)r^{2(p-1)},~~
R=\frac{2(p-1)}{p\Gamma(p)^2\sin(\pi p/2)^2},\n
R_{ab}=\frac{1}{2}Rg_{ab},~~~G_{ab}=0,~~~RiemSq=R^2,~~~RicciSq=\frac{1}{2}R^2.
\label{newcurv}
\ear
Like all two dimensional line elements
this solution obeys the vacuum Einstein equations.
For fixed constant $p$ it has constant curvature.
Using the method of signature constants \cite{mdrmd}
the signature of \ref{newd2} gives the same curvature in any combination.
Two dimensional line elements have been studies by Witten \cite{witten}.
Whether the line element \ref{newd2} is explicitly flat using the curvature \ref{genriemann}
leads to a calculation which has so far proved intractable.
\subsection{Fractional geodesics.}\label{fracgeodesics}%%%%%%%%%%%%%%%%%%%%%%%%%%%%%%%%%%%%%%%%
Instead of starting with \ref{genchristoffel} and \ref{genriemann} one could attempt a more
systematic approach by starting with a point particle lagrangian
\be
\Lg=\sqrt{-\dot{x}^2}\rightarrow
\left[-\left(\frac{dx^i}{d\tau}\right)^p\left(\frac{dx_i}{d\tau}\right)^p\right]^\frac{1}{2p},
\label{pointpart}
\ee
and then varying the fractional alteration to form a geodesic equation.
\ref{pointpart} can also be thought of as the integral form of an element of arc,
compare equation one \cite{chern}.
Once one has a geodesic equation one has a connection.
From the connection the commutation of covariant derivatives will give curvature.
The problem with this is that it is not clear what the right hand side of \ref{pointpart} means,
in principle it is not necessary to know explicity what \ref{pointpart} means only to know what the variation of \ref{pointpart} is,
this is unclear because terms such as $(\tf{dx^i}{d\tau})^p$ occur and there appears to be no way of defining them.
The same problems seem to occur starting from two other forms of the point particle lagrangian
such as the second-order form and the Hamiltonian form.
\section{Conclusion}\label{conc}%%%%%%%%%%%%%%%%%%%%%%%%%%%%%%%%%%%%%%%%%%%%%%%%%%%%%%%%%%%%%%%
It is possible to produce fractional generalizations of Newtonian mechanics and Friedman-Robertson-Walker
cosmology by replacing partial derivatives by fractional derivatives in familiar equations.
In cosmology the degree of fraction differentiation $p$ itself could be made a time dependent variable,
although this was not looked at here.
The fractional derivative results are what might have been anticipated:  however for consistency one should start
with fractional derivatives at the first step and this leads to the subject of fractional derivative geometry.
What fractional derivative geometry should look like is unclear,  here a first attempt was made at
guessing what curvature and line elements might look like in two dimensions.
By chance this lead to a line element depending on the Gamma-function.
%%%%%%%%%%%%%%%%%%%%%%%%%%%%%%%%%%%%%%%%%%%%%%%%%%%%%%%%%%%%%%%%%%%%%%%%%%%%%%%%%%%%%%%%%%%%%%

\begin{figure}
\includegraphics[height=5in]{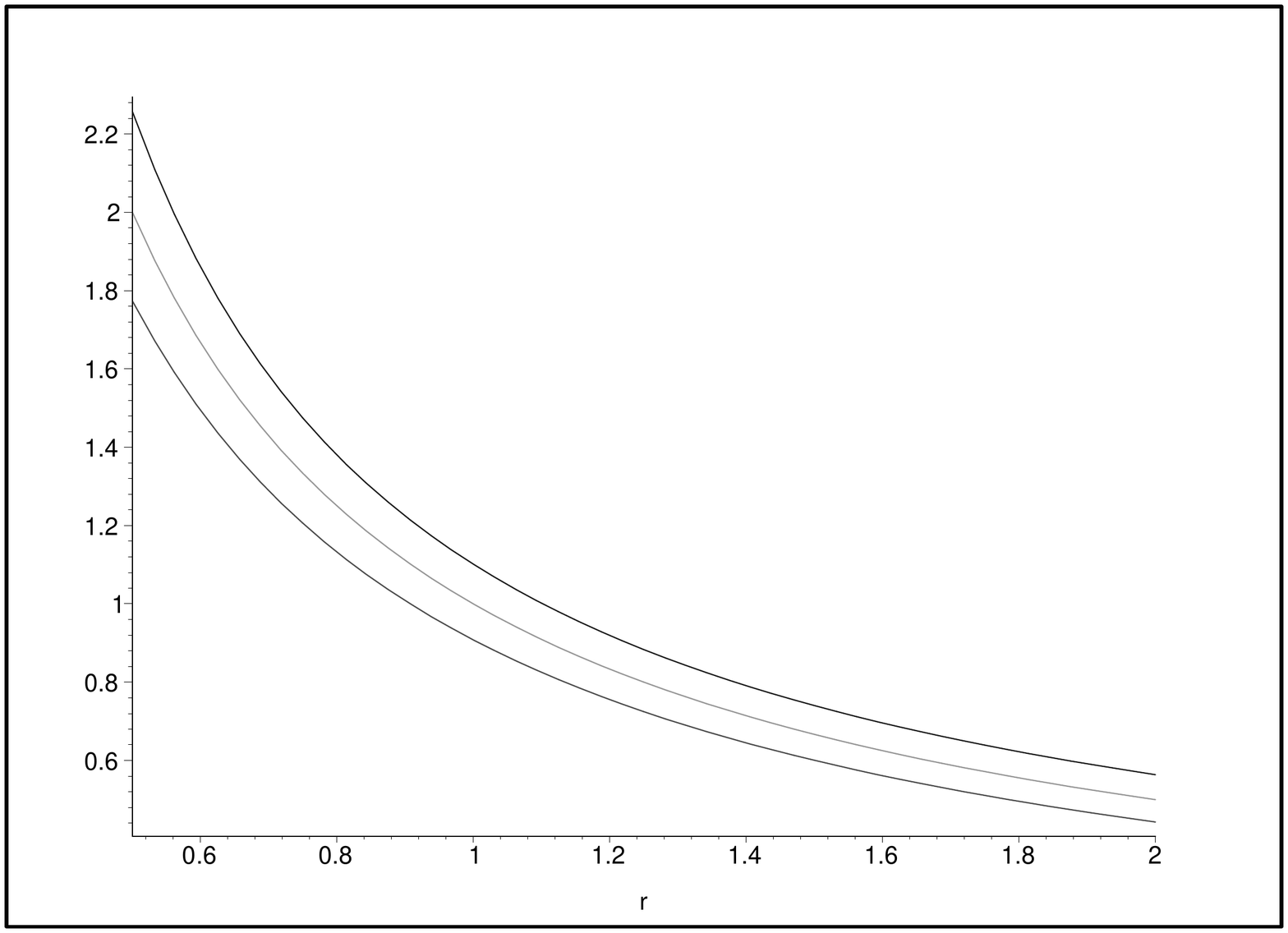}
\caption{lowest line$=2.10^4\exp(-r^{0.1}/0.1)$,
         middle line$=1/r$,
         top line$=5.10^{-5}\exp(r^{-0.1}/0.1)$}
\label{FigureOne}
\end{figure}
\begin{figure}
\includegraphics[height=5in]{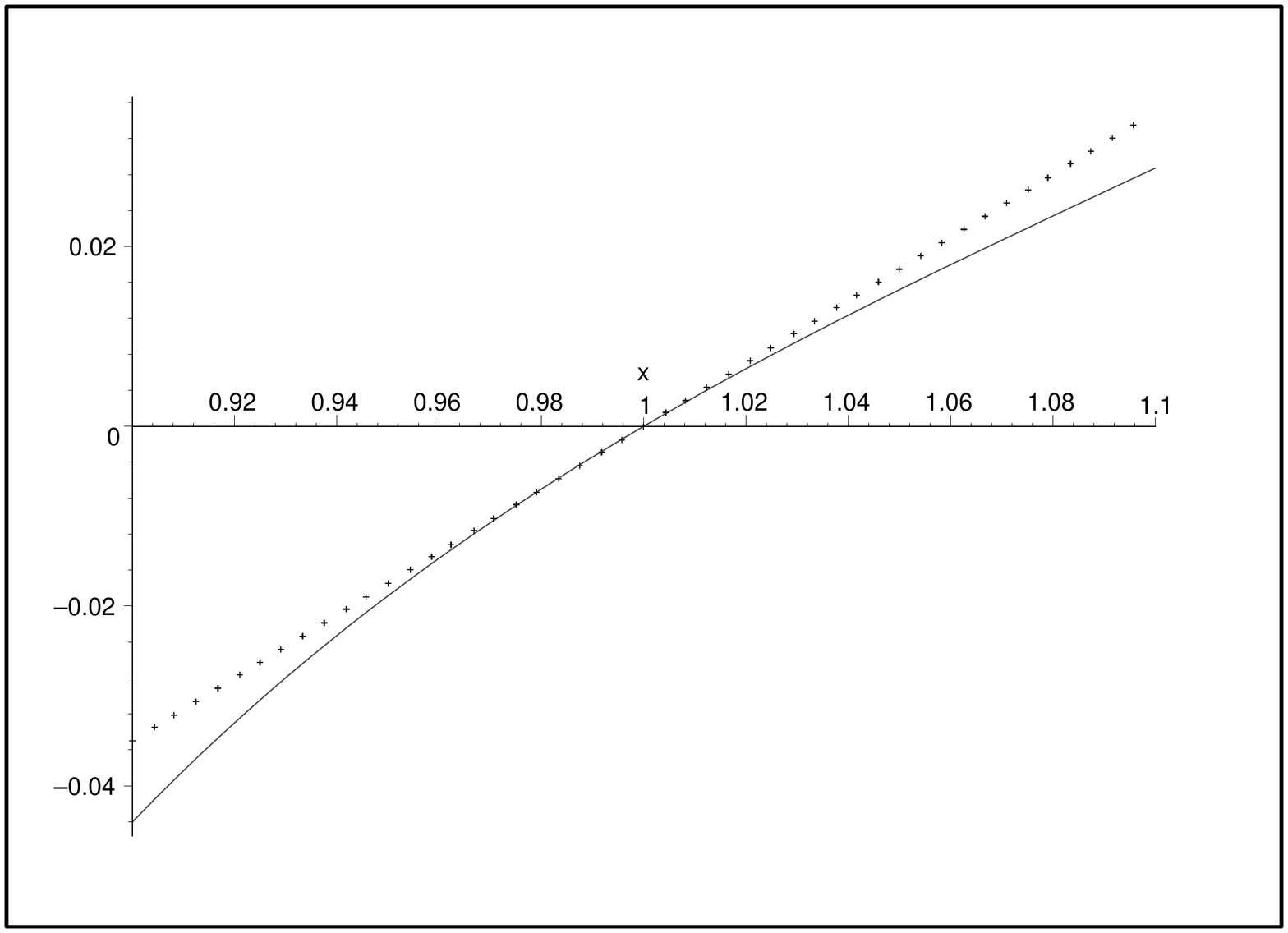}
\caption{Red line is $AC^2$,  blue points are $y=0.35(x-1)$}
\label{Compensating Pressure}
\end{figure}

\begin{thebibliography}{99}


\bibitem{BM}
Dumitru Baleanu \& Sami I. Muslih,
Lagrangian formulation of classical fields within Riemann-Liouville fractional derivatives,
\href{http://arXiv.org/abs/hep-th/0510071}
                      {\tt hep-th/0510071}

\bibitem{chern}
Shing-Shen Chern,
Finsler geometry is just Riemannian geometry without the quadratic restriction.
{\it Notices of the AMS}{\bf 43}(1996)959-963.

\bibitem{KST}
{\it Theory and Applications of Fractional Differential Equations},
by Anatoly A. Kilbas, Hari M. Srivastava \& Juan J. Trujillo,
Elsevier (2006).
\href{http://www.ams.org/mathscinet-getitem?mr=2007a:34002}
{Math.Rev. 2007a:34002}
ISBN 0-444-51832-0

\bibitem{laskin}
Nick Laskin,
Fractional Quantum Mechanics,
{\it Phys.Rev.E}{\bf 62}(2000)1315-1345,
\href{http://arXiv.org/abs/0811.1769}
                      {\tt 0811.1769}

\bibitem{MPL}
Peter Musgrave,  Dennis Pollney \& Kayll Lake,
(1996) GRTensorII,
\href{http://astro.queens.ca/~grtensor/}
     {http://astro.queens.ca/$\sim$grtensor/}

\bibitem{naber}
Mark Naber,
Time fractional Schr\"odinger equation,
{\it Journal of Mathematical Physics}{\bf 45}(2004)3339-3352.
[DOI: 10.1063/1.1769611]

\bibitem{podlubny}
Igor Podlubny,
Fractional Differential Equations,
(Mathematics in Science and Engineering, vol. 198),
Academic Press;
\href{http://www.ams.org/mathscinet-getitem?mr=99m:26009}
{Math.Rev. 99m:26009}
ISBN 0-12-558840-2

\bibitem{mdr39}
Mark D. Roberts,
Spacetime Exterior to a Star.
{\it International Journal of Theoretical Physics}
(2004)1-37.
\href{http://arXiv.org/abs/gr-qc/9811093}
                      {\tt gr-qc/9811093}

\bibitem{mdrmd}
Mark D. Roberts,
New Five Dimensional Spherical Vacuum Solutions.
\href{http://arXiv.org/abs/0901.2307}
                      {\tt 0901.2307}

\bibitem{witten}
Edward Witten,
String theory and black holes.
{\it Phys.Rev.D}{\bf 44}(1991)314-324,eq.22.

\bibitem{zavada}
Petr Zavada,
Relativistic wave equations with fractional derivatives and pseudodifferential operators,
{\it Journal of Applied Mathematics}{\bf 2}(2002)163-197.
\href{http://arXiv.org/abs/hep-th/0003126}
                      {\tt hep-th/0003126}

\end{thebibliography}
\end{document}